\documentclass[prl,aps,superscriptaddress,twocolumn,letter,nopacs,longbibliography]{revtex4-1}

\usepackage{array,multirow,graphicx,color,amsmath,amsfonts,enumerate,amsthm,amssymb,mathtools,enumitem,thmtools,hyperref,mathdots,enumitem,centernot,bm,soul,bbm,tikz,pgfplots,comment}
\usepackage[caption=false,subrefformat=parens,labelformat=empty]{subfig}
\usepackage{enumitem}
\usepackage[capitalise, noabbrev]{cleveref}
\usetikzlibrary{arrows}
\hypersetup{colorlinks=true,linkcolor=blue,citecolor=blue,urlcolor=blue}
\usepackage{braket}

\usepackage{soul}

\def\Re{\operatorname{Re}}

\newcommand{\abs}[1]{\left| {#1} \right|} 

\newcommand{\tr}[1]{\mathrm{Tr}\left[ #1 \right]}

\renewcommand{\v}[1]{\ensuremath{\boldsymbol #1}}

\definecolor{ppblue}{RGB}{46,117,182}
\definecolor{ppred}{RGB}{197, 90, 17}

\theoremstyle{plain}

\theoremstyle{definition}

\newcolumntype{C}[1]{>{\centering\arraybackslash}p{#1}}

\definecolor{tikzBlue}{rgb}{0.6941176470588235,0.7568627450980392,0.8588235294117647}
\definecolor{tikzOrange}{rgb}{0.9294117647058824,0.7647058823529411,0.49019607843137253}
\definecolor{tikzBlue2}{rgb}{0.462745098,0.504575163,0.57254902}
\definecolor{tikzOrange2}{rgb}{0.619607843,0.509803922,0.326797386}
\definecolor{tikzGray}{rgb}{0.7529411764705882,0.7529411764705882,0.7529411764705882}

\DeclareFontFamily{U}{mathb}{\hyphenchar\font45}
\DeclareFontShape{U}{mathb}{m}{n}{
	<-6> mathb5 <6-7> mathb6 <7-8> mathb7
	<8-9> mathb8 <9-10> mathb9
	<10-12> mathb10 <12-> mathb12
}{}
\DeclareSymbolFont{mathb}{U}{mathb}{m}{n}
\DeclareMathSymbol{\llcurly}{\mathrel}{mathb}{"CE}
\DeclareMathSymbol{\ggcurly}{\mathrel}{mathb}{"CF}

\begin{document}

\title{Projective measurements can probe non-classical work extraction and time-correlations}

\author{Santiago Hern\'{a}ndez-G\'{o}mez}
\email{shergom@mit.edu}
\altaffiliation{Current address: Research Laboratory of Electronics, Massachusetts Institute of Technology, Cambridge, MA 02139}
\affiliation{European Laboratory for Non-linear Spectroscopy (LENS), Universit\`a di Firenze, I-50019 Sesto Fiorentino, Italy}
\affiliation{Dipartimento di Fisica e Astronomia, Universit\`a di Firenze, I-50019, Sesto Fiorentino, Italy}
\affiliation{Istituto Nazionale di Ottica del Consiglio Nazionale delle Ricerche (CNR-INO), I-50019 Sesto Fiorentino, Italy}

\author{Stefano Gherardini}
\email{stefano.gherardini@ino.cnr.it}
\affiliation{Istituto Nazionale di Ottica del Consiglio Nazionale delle Ricerche (CNR-INO), Science Park, Basovizza, I-34149 Trieste, Italy}
\affiliation{European Laboratory for Non-linear Spectroscopy (LENS), Universit\`a di Firenze, I-50019 Sesto Fiorentino, Italy}

\author{Alessio Belenchia}
\affiliation{Institut f\"{u}r Theoretische Physik, Eberhard-Karls-Universit\"{a}t T\"{u}bingen, 72076 T\"{u}bingen, Germany}
\affiliation{Centre for Theoretical Atomic, Molecular and Optical Physics, School of Mathematics and Physics, Queen's University Belfast, Belfast BT7 1NN, United Kingdom}

\author{Matteo Lostaglio}
\email{lostaglio@protonmail.com}
\affiliation{Korteweg-de Vries Institute for Mathematics and QuSoft, University of Amsterdam, The Netherlands}

\author{Amikam Levy}
\affiliation{Department of Chemistry and Center for Quantum Entanglement Science and Technology, Bar-Ilan University, Ramat-Gan 52900, Israel}

\author{Nicole Fabbri}
\affiliation{European Laboratory for Non-linear Spectroscopy (LENS), Universit\`a di Firenze, I-50019 Sesto Fiorentino, Italy}
\affiliation{Istituto Nazionale di Ottica del Consiglio Nazionale delle Ricerche (CNR-INO), I-50019 Sesto Fiorentino, Italy}

\begin{abstract}
We demonstrate an experimental technique to  characterize genuinely nonclassical multi-time correlations using projective measurements with no ancillas. We implement the scheme in a nitrogen-vacancy center in diamond undergoing a unitary quantum work protocol. We reconstruct quantum-mechanical time correlations encoded in the Margenau-Hills quasiprobabilities.  
We observe work extraction peaks five times those of sequential projective energy measurement schemes and in violation of newly-derived stochastic bounds. We interpret the phenomenon via anomalous energy exchanges due to the underlying negativity of the quasiprobability distribution.
\end{abstract}

\maketitle

There is no unique way of defining a joint probability for the multi-time statistics of quantum mechanical quantities since even a single quantum observable does not always commute with itself at different times. Nevertheless, across the quantum sciences, multi-time fluctuations of quantum-mechanical quantities, especially energy, play a crucial role. 
Correlation functions between events at different times resemble joint probability distributions for the eigenvalues of the observables, but they are in general neither real nor positive. In fact, they are represented by \emph{quasiprobabilities}, akin to the well-known Wigner function in quantum optics~\cite{PhysRev.40.749}, but associated with a process rather than a state. The standard definition of a temporal correlation function between two events described by projectors $\Pi_i(0)$ and $\Pi_f(t)$,
\begin{equation}
    q^{\textrm{KD}}_{if} = \tr{\rho \Pi_i(0) \Pi_f(t)},
\end{equation}
\emph{coincides} with a 
Kirkwood-Dirac quasiprobability (KDQ)~\cite{kirkwood1933quantum, dirac1945analogy, ArvidssonShukurJPA2021}, and the same goes for multi-time extensions. 

As we survey in~\cite{companion_theory_paper}, the centrality of the KDQ and its real part, the Margenau-Hill quasiprobability (MHQ), has only recently come to be fully appreciated. 
These quasiprobabilities underpin analysis from perturbation theory~\cite{companion_theory_paper} to information scrambling~\cite{Yunger2017jarzynski,yunger2018quasiprobability}. Weak values~\cite{aharonov1988how,Dressel2014colloquium,companion_theory_paper} are conditional KDQ~\cite{yunger2018quasiprobability, companion_theory_paper} and generalized \emph{anomalous weak values}  are in one-to-one correspondence with non-classical (negative or complex) KDQ quasiprobabilities~\cite{Johansen04,KunjwalPRA2019}.  Negative values of the MHQ are linked to quantum metrological advantages in both local and postselected setups~\cite{Zhang2015precision,arvidsson2020quantum,Lostaglio2020certifying} and to power output advantages in quantum thermodynamics~\cite{Lostaglio2020certifying}.

Contrary to investigations of multi-time processes via sharp measurements~\cite{dorner2013extracting,BatalhaoPRL14,An15,Xiong18}, experimental investigations with quasiprobabilities have mostly focused on characterizing states rather than processes. This has a long tradition in quantum optics, where phase-space quasiprobabilities have been used extensively for tomographic scopes~\cite{leibfried1996experimental,breitenbach1995squeezed,dunn1995experimental,smithey1993measurement}. To our knowledge, experimental access to the KDQ and MHQ has been limited to weak measurement schemes and, even then, mostly to characterize states~\cite{lundeen2011direct,PhysRevLett.108.070402,thekkadath2016direct,kim2018direct,PhysRevLett.126.100403}. 

In such a context, the aim of our work is two-fold: First, pave the way to the experimental study of the role of non-commutativity in temporal correlations via the MHQ. We do so by providing the first experimental demonstration of a \emph{weak two-point measurement} (wTPM) scheme~\cite{johansen2007quantum,companion_theory_paper}, a protocol that -- in contrast to weak measurement schemes -- requires neither ancillae nor fine-tuned system-ancilla couplings. In fact, we reconstruct the back-reaction-free limit encapsulated by quasiprobabilities not by a weak measurement, but by linearly combining different projective measurement schemes in such a way that different back-reactions cancel. Conceptually our idea can be seen as a twist on probabilistic error cancellation techniques in quantum computing, where several noisy circuits are sampled from to reconstruct an ideal error-free limit~\cite{cai2022quantum}. This technique can be deployed in quantum thermodynamics beyond the scope of our study.

Second, we lay down the theoretical and experimental ground for the study of non-classical energetic processes via the MHQ. We measure work extraction peaks in a driven three-level system up to five times those of the TPM scheme~\cite{campisi2011colloquium} and in violation of a newly introduced stochastic bound. We explain this phenomenon by interpreting negative probabilities as non-classical pathways of a stochastic process. 
Remarkably, the data required to witness genuinely non-classical effects via violations of the stochastic work bound can be obtained from measurements routinely performed in TPM experiments.

We put this forward as a theoretical  and experimental framework to interpret the recently observed energetics of superconducting qubits experiments \cite{stevens2022energetics}, and more generally in quantum thermodynamics experiments showcasing genuinely non-classical features.

\textbf{Non-classicality.--}
The KDQ encodes temporal correlations between quantum observables and so does its real part, the MHQ. Here  we focus on the latter.
Given two observables $A(0)\equiv\sum_i a_i \Pi_i(0)$ and  $B(t)\equiv\sum_f b_f\Xi_f(t)$ in terms of their eigenvalues $a_i$, $b_f$ and their eigenprojectors $\Pi_i(0)$, $\Xi_f(t)$, and a quantum channel $\mathcal{E}$ describing the system dynamics in the time interval $[0,t]$, the MHQ reads as
\begin{equation}\label{eq:kd}
	q_{if}(\rho,t) = \Re\tr{\rho\Pi_i (0) \mathcal{E}^\dag(\Xi_f(t))},
\end{equation} 
where $\mathcal{E}^\dag$ denotes the adjoint of $\mathcal{E}$ and $\rho$ is the quantum state at $t=0$. The MHQ is a \emph{quasiprobability}, as it satisfies $\sum_{if} q_{if}=1$ and $q_{if} \in \mathbb{R}$. The marginals over $i$ ($f$) reproduce the quantum outcome statistics of a measurement of $B(t)$ carried out at time $t$ ($A(0)$, carried out at time $0$). The MHQ, being a two-time correlator~\footnote{In fact, it can be extended to multi-time correlators.}, encodes information about the \textit{process}, including its linear response and quantum currents \cite{HovhannisyanNJP2019, companion_theory_paper}. 

In our experiments $A(0)$ and $B(t)$ are the Hamiltonian at times $0$ and $t$ and the channel is a unitary work protocol $U$, i.e., $\mathcal{E}(\cdot) \equiv \mathcal{U}(\cdot) \equiv U (\cdot) U^\dagger$. The `unperturbed' work $\langle w \rangle_t := \tr{H(t) \rho(t)} - \tr{H(0) \rho(0)}$ can be obtained as the average
\begin{equation}
\label{eq:work}
    \langle w \rangle_t = \sum_{i,f} q_{if}(\rho,t) w_{if}
\end{equation}
with $w_{if} = E_f(t) - E_i(0)$. Clearly, the same framework applies to the study of temporal correlations beyond work processes.

The quantum process has a stochastic (classical) interpretation when $q_{if} \geq 0$ for all $i,f$. In fact, for fixed $i,f$, commutativity implies positivity: if (a) $[\rho,\Pi_i]=0$ or (b) \mbox{$[\Pi_i, \mathcal{E}^\dag(\Xi_f)]=0$} or (c) $[\mathcal{E}^\dag(\Xi_f), \rho]=0$, then $q_{if} \geq 0$~\cite{companion_theory_paper}. The converse does not hold, i.e., negativity is a \emph{stronger} property than non-commutativity ~\cite{ArvidssonShukurJPA2021}. In cases (a-b), $q_{if}$ reduces to the TPM probability of observing outcomes $i$ followed by $f$ in a sequential projective measurement of the observables $A(0)$ and $B(t)$, with the intermediate evolution $\mathcal{E}$.

Negative values of $q_{if}$ indicate non-classicality in the temporal correlations. These are associated with proofs of contextuality~\cite{PuseyPRL2014, KunjwalPRA2019} and correspond to elementary non-classical processes. For work protocols, crucially an \emph{anomalous excitation} process $E_f > E_i$ (classically associated to work done, not extracted!) occurring with `negative probability' $q_{if}<0$ is equivalent to a \emph{classical de-excitation} process occurring with probability $|q_{if}|$, and hence contributes to the extracted work.

The non-classicality of the MHQ is defined via \textit{the negativity functional}~\cite{ArvidssonShukurJPA2021,alonso2019out, companion_theory_paper}~\footnote{The same expression holds true for the full KDQ, but it encodes not only the possibility of negative elements but also of imaginary ones.}
\begin{equation}\label{eq:negativity}
   \aleph \equiv -1+\sum_{i,f}|q_{if}(\rho,t)|\,.
\end{equation}
For work extraction from pure states, we prove in the supplemental material~\cite{SM} an upper bound on the extracted work $W_{\rm ext}$ that holds whenever a stochastic interpretation is possible, i.e., \mbox{$\aleph=0$}:
\begin{equation}\label{eq:bound_work_extraction}
    W_{\rm ext} \equiv 
    - \langle w\rangle_t \leq \sum_{i,f \,\, \text{s.t.} \,\, w_{if}>0}w_{if} \sqrt{p^{\rm TPM}_{if}p^{\rm END}_f}\,,
\end{equation}
where $p^{\mathrm{TPM}}_{if} \equiv p_i \tr{\mathcal{U}(\Pi_i)\Xi_f(t)}$, with $p_i = \tr{\rho\Pi_i}$, are the joint probabilities from the TPM scheme, and $p^{\mathrm{END}}_{f} \equiv \tr{\mathcal{U}(\rho)\Xi_f(t)}$ is the END-time energy measurement probability~\cite{gherardini2021end}. Standard TPM experiments satisfy this inequality. Hence, its violations indicate work extraction peaks above TPM that can only occur because negativity is at play. We will look for these peaks in the experimental data by optimizing the negativity of anomalous excitation processes within the experimentally achievable parameters.

\textbf{Measurement scheme.--}
We present an experimental implementation of the wTPM measurement scheme~\cite{johansen2007quantum}, a non-selective (NS) $2$-outcome projective measurement that checks whether the initial energy is $E_i$ or NOT $E_i$, followed by unitary evolution and a projective measurement of the final Hamiltonian. The wTPM joint probabilities read
\begin{equation}
    p^{\mathrm{wTPM}}_{if} \equiv \tr{\mathcal{U}(\rho_{\mathrm{NS},i})\Xi_f(t)},
\end{equation}
where $\rho_{\mathrm{NS},i} = p_i \rho_i + (1-p_i) \overline{\rho}_i$, $\rho_i = \Pi_{i}\rho\Pi_i/p_i$, and $\overline{\rho}_i = (\mathbb{I}-\Pi_i) \rho(\mathbb{I}-\Pi_i)/(1-p_i)$. The state $\rho_{\mathrm{NS},i}$ can be obtained by performing non-selective projective measurements with projectors $\Pi_i$ and $\mathbb{I}-\Pi_i$ or, equivalently, by the preparation of the states $\rho_i$ and $\overline{\rho}_i$ with the corresponding probabilities (as in our experiments). 

This joint probability is related to the MHQ~\cite{companion_theory_paper,johansen2007quantum} by
\begin{equation} \label{eq:MHQ}
    q_{if}= p^{\mathrm{TPM}}_{if} - \frac{1}{2}\left(p^{\mathrm{wTPM}}_{if} - p^{\mathrm{END}}_{f}\right),
\end{equation}
i.e., the MHQ is given by three distinct contributions~\footnote{ For the special case of a two-level quantum system, the TPM and the end-point measurement suffice to completely characterize $q_{if}$.} that stem from applying in three separate sets of runs the wTPM protocol, the TPM and END schemes~\footnote{The central operator equality that allows operating the above `probabilistic error cancellation' scheme is the following:
\begin{equation*}
    \Pi_i = \Pi_i \rho \Pi_i - 1/2 [(I-\Pi_i) \rho (I-\Pi_i) + \Pi_i \rho \Pi_i] + 
    \rho / 2 \,,
\end{equation*}
where each term on the right-hand-side of the equality can be associated with a different projective measurement scheme.
}.

\textbf{Experimental setting.}--
We use as a quantum system the electronic spin of an NV center in bulk diamond at room temperature. NV centers are defects in a diamond lattice with an orbital ground state that is a spin triplet $S=1$. The degeneracy in the spin quantum number $m_S$ is lifted due to the zero field splitting and to the presence of an external bias field aligned with the spin quantization axis. The NV spin qutrit can be optically initialized into $m_S=0$ ($\ket{0}$) by illuminating the defect with a green laser~\cite{Harrison04}. 
Moreover, the spin state can be read out by detecting the photoluminescence (PL) after a laser illumination, as the PL intensity depends on the spin projection $m_S$~\cite{Jelezko04,Doherty13}.
In addition, on-resonance microwave fields are used to coherently drive the spin, with coherence times up to milliseconds (at room temperature)~\cite{Balasubramanian09,Bar-Gill13}. By virtue of these properties, NV centers are broadly used for quantum technologies, such as quantum sensing~\cite{Rondin14,Degen17,HernandezGomez21Frontiers}, quantum information~\cite{Aharonovich14,Bradley19} and, recently, for quantum thermodynamics~\cite{HernandezGomezPRR20,HernandezGomez21,HernandezGomez2021nonthermal}.

A time-varying Hamiltonian is implemented by coherently driving the NV spin with a microwave field with phase changing in time. More specifically, the spin qutrit is driven by a bi-chromatic microwave field on-resonance with both the transitions $\ket{0}\leftrightarrow\ket{-1}$ and $\ket{0}\leftrightarrow\ket{+1}$. In the microwave rotating frame, the Hamiltonian of the system (after the rotating wave approximation) is 
\begin{align}
H(t) = \; &\Omega_1\left( S_{x1}\cos\phi_1 t +S_{y1}\sin\phi_1 t \right) + \notag\\ 
& \Omega_2\left( S_{x2}\cos\phi_2 t -S_{y2}\sin\phi_2 t  \right),
\label{eq:Hamiltonian}
\end{align} 
where $\hbar=1$, $S_{\alpha}$ are the spin operators defined in terms of the Gell-Mann matrices, and the Hamiltonian amplitude $\Omega_1$~($\Omega_2$) and rate of phase increase $\phi_1$~($\phi_2$) correspond respectively to the Rabi frequency and the phase of the driving field for the transition $\ket{0}\leftrightarrow\ket{+1}$~($\ket{0}\leftrightarrow\ket{-1}$), as detailed in~\cite{SM}. 
See also~\cite{SM} for more details on the energy level structure and the driving fields. To simplify the measurements in the time-varying energy eigenbasis, we remove the time dependency on one of the Hamiltonian eigenstates by setting  $\Omega_1=\Omega_2=\Omega$ and $\phi_1=\phi_2=\phi$. 
The eigenstates of the Hamiltonian~\eqref{eq:Hamiltonian} are: $\ket{E_{\pm}(t)} = \frac{1}{2}(\ket{1} \pm \frac{1}{\sqrt{2}}e^{i\phi t}\ket{0} + \ket{-1})$, and $\ket{E_{0}(t)} = \frac{1}{\sqrt{2}}\left(\ket{-1}-\ket{1}\right)$, with eigenvalues $E_{\pm}=\pm\Omega$ and $E_{0}=0$. Thus, the projectors of interest are $\Xi_k (t)= \ket{E_{k}(t)}\!\bra{E_{k}(t)}$ and $\Pi_k(0) = \ket{E_{k}(0)}\!\bra{E_{k}(0)}$ for $k=+, 0,-$. 
Note that $[H(0),H(t)]\neq 0$ for $t\phi/\pi \not\in\mathbb{Z}$. Hence, the system energy changes during the unitary evolution under $H(t)$, meaning that work is exchanged between the NV spin and the microwave field.

\textbf{Work quasiprobabilities.--} 
The MHQ plays a role in thermodynamics, where the characterization of work, heat, and internal energy fluctuations in quantum processes calls for novel tools to account for exquisitely 
quantum effects. The seminal TPM scheme~\cite{TalknerPRE2007,campisi2011colloquium,RevModPhys.81.1665} is unable to capture non-commutativity \cite{lostaglio2018quantum}. This motivated the use of the MHQ to characterize non-classical work fluctuations~\cite{allahverdyan2014nonequilibrium} and anomalous heat exchanges due to quantum correlations~\cite{levy2020quasiprobability}. However, experimental realizations remained limited, due to the challenge of adapting TPM experiments to access work quasiprobabilities. Here, we experimentally reconstruct the MH work quasiprobability on the NV center using only projective energy measurements and pure state preparations. This paves the way for a range of other TPM experiments to adopt the same strategy.

\begin{figure}[t!]
    \centering
    \includegraphics[width=\columnwidth]{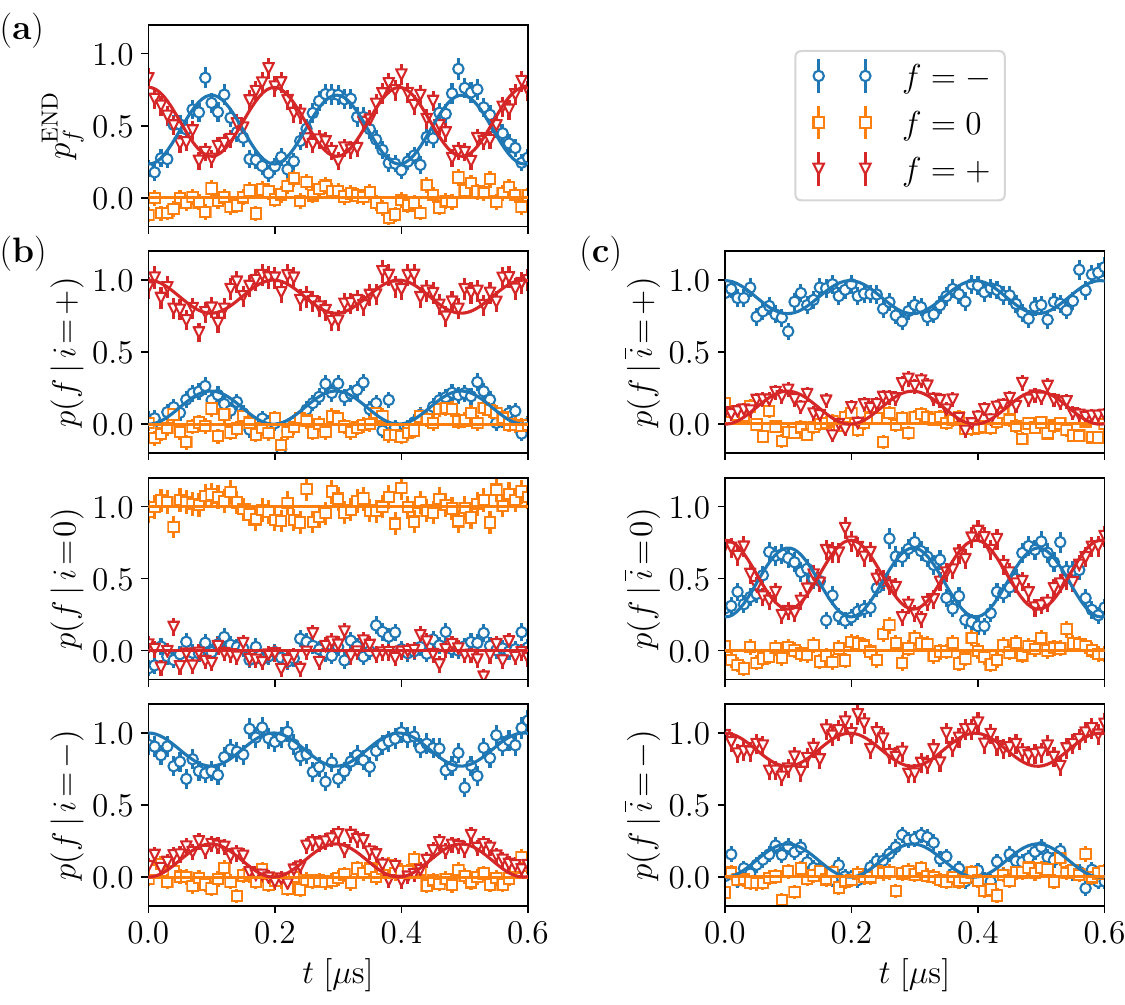} \caption{
    Experimental results from the END scheme [panel (a)], and from the measurements of the conditional probabilities obtained by initializing the qutrit in the states $\rho_i$~[panels (b)] and $\overline{\rho}_i$~[panels (c)]. 
    The solid lines denote the simulations using Eq.~\eqref{eq:Hamiltonian} with $\Omega=(2\pi)2.219$~MHz and $\phi = 1.09\,\Omega$. 
    Note that the data for f=0 are always constant. This is a consequence of setting $\phi_1=\phi_2=\phi$. In such case, the interaction between the NV center and the two microwave fields corresponds to a Stimulated Raman Adiabatic Passage (STIRAP) in the two-photon resonance condition~\cite{Vitanov17} (see also~\cite{SM}).
    The data outside the interval $[0,1]$ is originated by photon shot noise during the PL read-out, hence affecting the PL normalization (see text).
    }
    \label{fig:measurements_END_and_conditional}
\end{figure}

\begin{figure*}[t!]
\centering
\includegraphics[width=\textwidth]{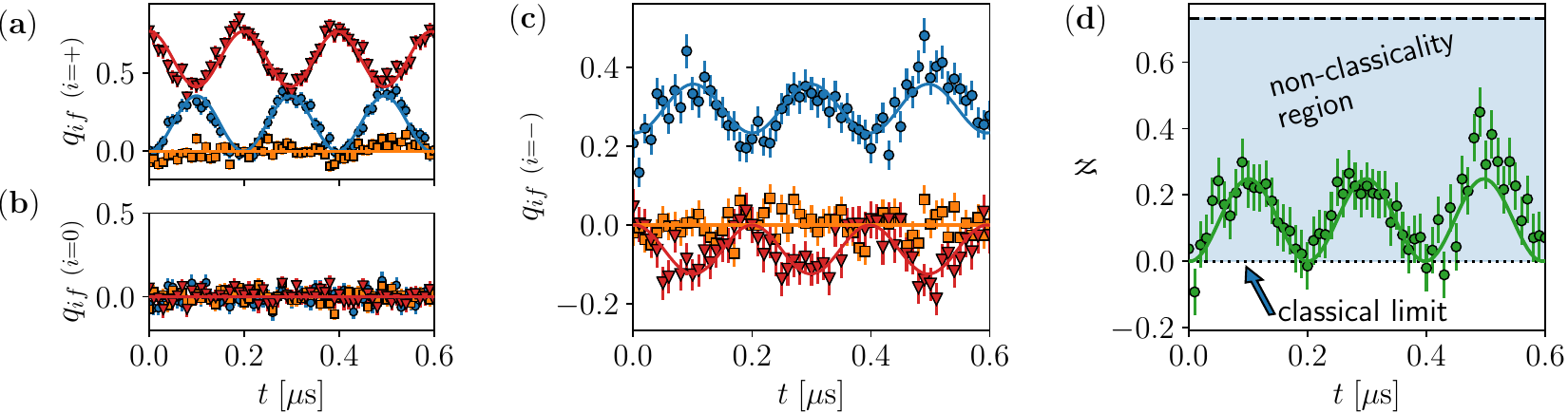} 
\caption{
(a-c)~Measured MHQs $q_{if}$ as a function of time.  
The blue circles correspond to $f={-}$, the orange squares to $f={0}$, and red triangles to $f={+}$. The solid line represents the simulated data, while the dashed black line corresponds to $\sum_f \abs{ q_{if}}$. 
Note that only $q_{-+}$~[panel (c)] exhibits negative values. 
(d)~Experimental measurements (green circles) of the negativity~[Eq.~\eqref{eq:negativity}] as a function of time (solid green line represents the simulated data). For almost all the interaction time $t$ the negativity is larger than zero, hence overcoming the classical limit (dotted line) and entering into the non-classical region (blue area). This region is bounded from above by $\sqrt{d}-1$ (dashed line), where $d=3$ is the dimension of the system's Hilbert space~\cite{companion_theory_paper}.}
\label{fig:MHQ_and_negativity}
\end{figure*}

We take the initial state to be pure $\rho = \ket{\xi}\!\bra{\xi}$~\footnote{The initial state $\ket{\xi}$ is chosen, over an ensemble of $1000$ initial pure random states~\cite{SM}, as the one that minimizes $q_{-+}$ when considering the Hamiltonian~\eqref{eq:Hamiltonian} with $\Omega\simeq(2\pi)2.2$~MHz and $\phi \simeq 1.09\,\Omega$. This results in $\ket{\xi}\equiv \sum_i \sqrt{p_i} e^{2 \pi j a_i}\ket{E_{i}(0)}$, with $j^2=-1$, $p_i = 0.7654, 0.0009, 0.2338$ and $a_i = 0.0073,0.2787,0.0002$ for $i=+,0,-$ respectively.}. One can therefore reconstruct $p^{\mathrm{END}}_{f}$, $p^{\mathrm{TPM}}_{if}$, and $p^{\mathrm{wTPM}}_{if}$ 
by measuring a set of conditional probabilities of the form 
\begin{equation}\label{eq:p_f_cond_psi}
p(f|\psi) \equiv \tr{\mathcal{U}(\ket{\psi}\!\!\bra{\psi})\Xi_f(t)},
\end{equation}
where $\ket{\psi}$ depends on the scheme that we want to implement. 
We directly measure $p^{\mathrm{END}}_{f} = p(f|\xi)$ for the END scheme. Instead, for the TPM and wTPM schemes we measure the conditional probabilities $p(f|i)$ and $p(f|\overline{i})$ by initializing the quantum states $\rho_i$ and $\overline{\rho}_i$, respectively, and we combine them as $p^{\mathrm{TPM}}_{if} = p_{i}\,p(f|i)$ and $p^{\mathrm{wTPM}}_{if} = p_{i}p(f|i) + (1-p_i) p(f|\overline{i})$ (we recall $p_i = {\rm Tr}[\rho\Pi_i]$).

The results for the measurements of $p^{\mathrm{END}}_{f}$, $p(f|i)$ and $p(f|\overline{i})$ are shown in Fig.~\ref{fig:measurements_END_and_conditional}. The protocol to measure these conditional probabilities is based on our previous works~\cite{HernandezGomezPRR20,HernandezGomez21,HernandezGomez2021nonthermal}; however, the full description of the protocol is also included in~\cite{SM}. The main idea is the following: First, the qutrit is prepared into the pure state $\ket{\psi} = \ket{\xi}$, $\ket{i}$ or $\ket{\overline{i}}$, depending on the measurement scheme (END, TPM, wTPM). Then, it evolves under the time-varying Hamiltonian $H(t)$ in the time interval $[0,t]$. 
At the end of the protocol, we optically read out the probability that the energy of the system is $E_f(t)$, i.e., $p(f|\psi)$. As mentioned before, the PL intensity (averaged over $\sim10^6$ repetitions of the experiment) encodes information about the spin state. Hence, by normalizing the average PL with respect to reference PL levels we obtain $p(f|\psi)$~\cite{SM}. 
Note that the optical read-out is destructive, hence, for each given initial state, we perform independent experiments for each value of $t$ and for each of the three Hamiltonian projectors $\Xi_f(t)$.

We can now obtain the MHQ work distributions at each $t$ by combining the results of all the previous measurements as dictated by Eq.~\eqref{eq:MHQ}. The results are shown in Fig.~\ref{fig:MHQ_and_negativity}a-c. 
From Eq.~\eqref{eq:negativity} the non-classicality is quantified by the negativity of the measured work distribution. 
Its experimental values are plotted in Fig.~\ref{fig:MHQ_and_negativity}d.

\begin{figure}
\centering
\includegraphics[width=1\columnwidth]{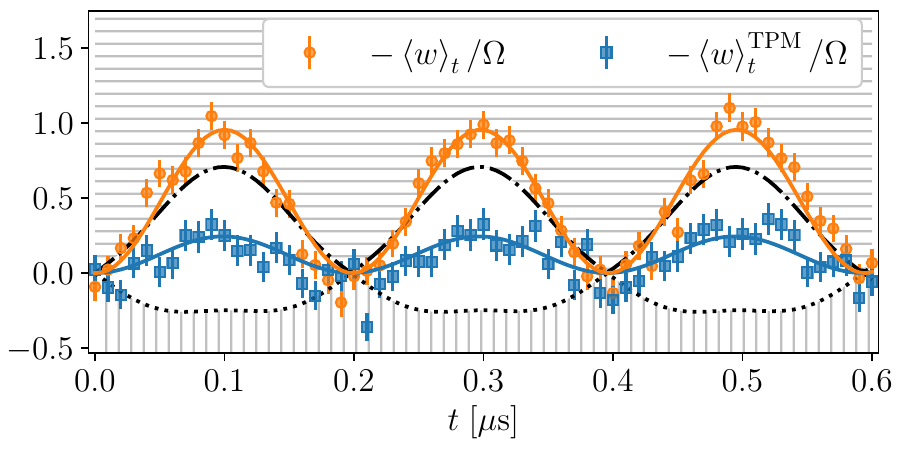}
\caption{
Average unperturbed extracted work $W_{\rm ext}=-\langle w\rangle_t = \sum_{i}p_i E_{i}(0) - \sum_{f}p^{\mathrm{END}}_{f} E_{f}(t)$, and average extracted work in the TPM scheme, $-\langle w\rangle_t^{\rm TPM} = -\sum_{i,f}p^{\mathrm{TPM}}_{if} w_{if}$. The striped region on top (bottom) indicates violations of the stochastic work extraction (injection) bounds, achievable only if $\aleph\neq0$~[see Eq.~\eqref{eq:bound_work_extraction} and \cite{SM}].
}
\label{fig:mean_work_bis}
\end{figure}

\textbf{Work extraction.--}
Let us focus now on the thermodynamics of the driven qutrit. 
In Fig.~\ref{fig:mean_work_bis} we compare the experimental data for the average extracted work in the TPM scheme, $-\langle w\rangle_t^{\rm TPM}$, 
with the unperturbed extracted work $W_{\rm ext}=-\langle w\rangle_t$. In our experiment, the TPM (projective measurements) reduces the efficiency of the work extraction process. 
Comparing Fig.~\ref{fig:mean_work_bis} and Fig.~\ref{fig:MHQ_and_negativity} we observe that peaks in the average work coincide with peaks in negativity (non-classical process). What is more, 
Fig.~\ref{fig:mean_work_bis} shows that the stochastic bound of Eq.~\eqref{eq:bound_work_extraction} for work extraction is violated, showing that these peaks are high enough that they can \emph{only} occur when $q_{if}$ turns negative. The bound in Eq.~\eqref{eq:bound_work_extraction} is a powerful tool for \emph{witnessing} non-classicality, as it relies only on the combination of the TPM statistics and END-time energy measurements.

A physical interpretation of the non-classical work can be given by noting that, in our experiments, negativity is concentrated in the anomalous excitation processes -- negativity of the MHQ distribution is associated with the largest exciting transition, $w_{-+} = 2\Omega$ [Fig.~\ref{fig:MHQ_and_negativity}a-c]. 
Classically this transition contributes to work done but quantumly it enhances work extraction. This negativity is destroyed in the TPM scheme, resulting in decreased work extraction. Theoretical considerations often focus on the total negativity $\aleph$, but from a thermodynamic point of view, our experiments indicate that it is the \emph{distribution of negativity} among the outcomes that plays a crucial role. In fact, numerical simulations~\cite{SM} show that our experimental conditions are both close to minimizing $q_{-+}$ as well as maximizing the extracted work.

\textbf{Conclusions.}--
We presented the first experimental implementation of a wTPM scheme, reconstructing the Margenau-Hills work quasi-distribution for a spin qutrit. Our platform of choice has been an NV center in diamond driven by a microwave field acting as a work reservoir.

Our experiment demonstrates how to reconstruct genuinely non-classical effects in a work process using projective measurements only, without the need for ancillae and fine-tuned couplings, with similar resource requirements as the TPM. 

Furthermore, we found that peaks in the work extraction are associated with peaks in non-classicality in the form of negativity. In fact, the height of the observed work peaks was such that they can only be explained by the presence of negativity. We proved this by introducing a new stochastic bound for work processes [Eq.~\eqref{eq:bound_work_extraction}] that allows inferring negativity without even implementing the wTPM measurement scheme. Thus, we witness non-classicality with minimal adjustments to TPM experiments already in place.
Since the TPM scheme has been implemented in a variety of platforms~\cite{An15,Xiong18,HernandezGomezPRR20,HernandezGomez21,HernandezGomez2021nonthermal}, we expect our demonstration will herald further experimental studies on such set-ups.

We gave a general interpretation of the nonclassical work phenomenon as an anomalous energy process, where negativity flips the conventional directionality of selected stochastic transitions, transforming contributions to work done into contributions to work extracted -- leading to the observed peaks. We highlighted the thermodynamic relevance of the `negativity distribution', beyond usual considerations of total negativity. 
The enhanced work extraction resembles recently reported anomalous energy exchanges in superconducting qubit systems~\cite{stevens2022energetics,maffei2022anomalous}. We believe our work will provide a suitable interpretative and experimental framework for that setup, but we leave this study for future work.

Finally, our experiment can provide access to the multi-time correlations of a driven unitary dynamics. This suggests an alternative path to witnessing relevant process properties, such as scrambling~\cite{alonso2019out}, by combining independent experiments that involve projective measurements only and no ancillae. This is particularly relevant in view of the results in~\cite{CampisiPRE2017thermodynamics} linking, in the case of diagonal initial states, a TPM characteristic function to the out-of-order-correlations (OTOC) witnessing information scrambling. These results can be generalized beyond the diagonal case when access to the MHQ is granted. The potential use of wTPM schemes to extract information from the dynamics of many-body systems is another future line of research suggested by our work.

\begin{acknowledgments}
\emph{Acknowledgments.} We gratefully thank F. Poggiali for critical reading of the manuscript. S.H.G. acknowledges the financial support from CNR-FOE-LENS-2020. S.G. acknowledges The Blanceflor Foundation for financial support through the project ``The theRmodynamics behInd thE meaSuremenT postulate of quantum mEchanics (TRIESTE)''. A.B. acknowledges support from the Deutsche Forschungsgemeinschaft (DFG, German Research Foundation) project number BR 5221/4-1. A.L. acknowledges support from the Israel Science Foundation (Grant No. 1364/21). The work was also supported by the European Commission under GA n. 101070546--MUQUABIS.
\end{acknowledgments}

\bibliography{quantum.bib}

\newpage
\widetext

\clearpage

\begin{center}
\textbf{\large Supplemental Material: Projective measurements can probe non-classical work extraction and time-correlations}
\end{center}

\setcounter{equation}{0}

\setcounter{figure}{0}
\setcounter{table}{0}
\setcounter{page}{1}
\makeatletter
\renewcommand{\theequation}{S\arabic{equation}}
\renewcommand{\thefigure}{S\arabic{figure}}

\section*{Details on the experimental setup}

As described in the main text, the three level system realized for our experiments is based on the spin triplet $S=1$ of the orbital ground state of an NV center, with Hamiltonian \begin{equation}
    \mathcal{H}_{\mathrm{NV}} = \Delta S_z^2 + \gamma_e B S_z \,,
\end{equation}
where $\Delta =2.87 {\rm GHz}$ is the zero-field-splitting, $\gamma_e$ denotes the electron gyromagnetic ratio, and $B$ is a bias magnetic field aligned with the NV quantization axis $z$ (determined by the orientation of the NV defect in the diamond lattice). 

\begin{figure}[h!]
\centering
\includegraphics[width=0.75\columnwidth]{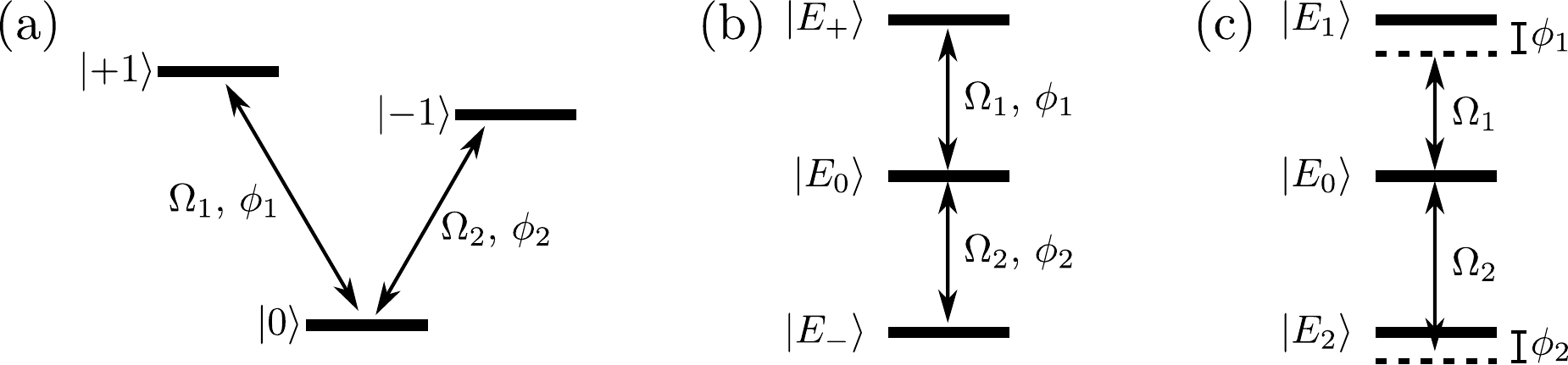}
\caption{
Energy diagram for the NV center spin triplet. (a) In the laboratory frame according to the Hamiltonian $\mathcal{H}(t)$ in Eq.~\eqref{eq:Hamiltonian_mathcal}.
(b) In the MW rotating frame, as described by the Hamiltonian $H(t)$ in Eq~\eqref{eq:Hamiltonian_bis} (or Eq.~(8) in the main text).
(c) In a rotating frame where the Hamiltonian is effectively time-independent, as in Eq.~\eqref{eq:Hamiltonian_tilde}. It is worth noting that in frame (c), the time-varying phase of each MW field can be reinterpreted as detuning. 
}
\label{fig:NV_E_levels}
\end{figure}

The spin triplet is driven by two continuous on-resonance microwave (MW) fields addressing the transitions $\ket{0}\leftrightarrow\ket{-1}$ and $\ket{0}\leftrightarrow\ket{+1}$. Hence, overall the spin dynamics can be described by the Hamiltonian 
\begin{equation}
\mathcal{H}(t) = \mathcal{H}_{\mathrm{NV}} +  \left( \Omega_1 \cos(\omega_{+1} t + \varphi_1(t) ) \ket{+1}\!\bra{0} + \Omega_2 \cos(\omega_{-1} t + \varphi_2(t)) \ket{-1}\!\bra{0} + \mathrm{h.c.} \right) , 
\label{eq:Hamiltonian_mathcal}
\end{equation}
where $\Omega_1$ and $\Omega_2$ are the Rabi frequencies for the transitions $\ket{0}\leftrightarrow\ket{+1}$ and $\ket{0}\leftrightarrow\ket{-1}$, respectively. In addition, $\omega_{\pm1} = \Delta \pm \gamma_e B$ denote the frequencies, and $\varphi_1(t)$ and $\varphi_2(t)$ are the time-varying phases of the MW fields. The energy level structure of the qutrit and its interaction with the MW fields is depicted in Fig.~\ref{fig:NV_E_levels}(a). 

In the microwave rotating frame (defined by the unitary transformation $V = \exp[j t\allowbreak (\omega_{+1} \ket{+1}\!\bra{+1} + \omega_{-1} \ket{-1}\!\bra{-1} )]$, with $j^2=-1$) and after applying the rotating wave approximation, the Hamiltonian $\mathcal{H}(t)$ reads as
\begin{equation}
H(t) = \Omega_1\left( S_{x1}\cos\varphi_1(t) +S_{y1}\sin\varphi_1(t) \right) +  \Omega_2\left( S_{x2}\cos\varphi_2(t) +S_{y2}\sin\varphi_2(t)  \right).
\label{eq:Hamiltonian_bis}
\end{equation}
The Hamiltonian (\ref{eq:Hamiltonian_bis}) is defined in terms of the spin operators $S_{x1} = \frac{1}{\sqrt{2}}\lambda_1$, $S_{y1} = \frac{1}{\sqrt{2}}\lambda_2$, $S_{x2} = \frac{1}{\sqrt{2}}\lambda_6$, $S_{y2} = \frac{1}{\sqrt{2}}\lambda_7$, where
$\lambda_i$ are the Gell-Mann matrices:
\begin{equation}
    \lambda_1 = \begin{pmatrix}
    0 & 1 & 0 \\
    1 & 0 & 0 \\
    0 & 0 & 0
    \end{pmatrix} \;;\;
    \lambda_2 = \begin{pmatrix}
    0 & -i & 0 \\
    i & 0 & 0 \\
    0 & 0 & 0
    \end{pmatrix} \;;\;
    \lambda_6 = \begin{pmatrix}
    0 & 0 & 0 \\
    0 & 0 & 1 \\
    0 & 1 & 0
    \end{pmatrix} \;;\;
    \lambda_7 = \begin{pmatrix}
    0 & 0 & 0 \\
    0 & 0 & -i \\
    0 & i & 0
    \end{pmatrix}.
\end{equation}
In our experiments, we select the time-varying phases so that they change linearly in time, i.e., $\varphi_1(t) = \phi_1 t$ and $\varphi_2(t) = -\phi_2 t$. 
Therefore, Eq.~(8) in the main text is recovered from Eq.~\eqref{eq:Hamiltonian_bis}. 
The energy level structure in the MW rotating frame is sketched in Fig.~\ref{fig:NV_E_levels}(b). 

Finally, let us note that we can describe the system dynamics in a different rotating frame for which the Hamiltonian is time independent. In fact, instead of the MW rotating frame, we can express Eq.~\eqref{eq:Hamiltonian_mathcal} by transforming it in the rotating frame determined by the unitary transformation $V=\exp\left[j t \left((\omega_{+1}+\phi_1) \ket{+1}\!\bra{+1} + (\omega_{-1} + \phi_2) \ket{-1}\!\bra{-1} \right)\right]$.  
In this new frame and after the rotating wave approximation, the time-independent Hamiltonian is:
\begin{equation}
\tilde{H} = \Omega_1 S_{x1} - \phi_1 S_{z1} + \Omega_2 S_{x2} + \phi_2 S_{z2} \,, 
\label{eq:Hamiltonian_tilde}
\end{equation}
where $S_{z1} = \ket{+1}\!\bra{+1}$ and $S_{z2} = -\ket{-1}\!\bra{-1}$. 
The diagram of the energy level structure for this new rotating frame is shown in Fig.~\ref{fig:NV_E_levels}(c). Observe that this diagram represents a Stimulated Raman Adiabatic Passage (STIRAP) experiment for a three-level ladder scheme~\cite{Vitanov17}.

\section*{Protocol to measure conditional probabilities}

\begin{figure}[h!]
\centering
\includegraphics[width=0.55\columnwidth]{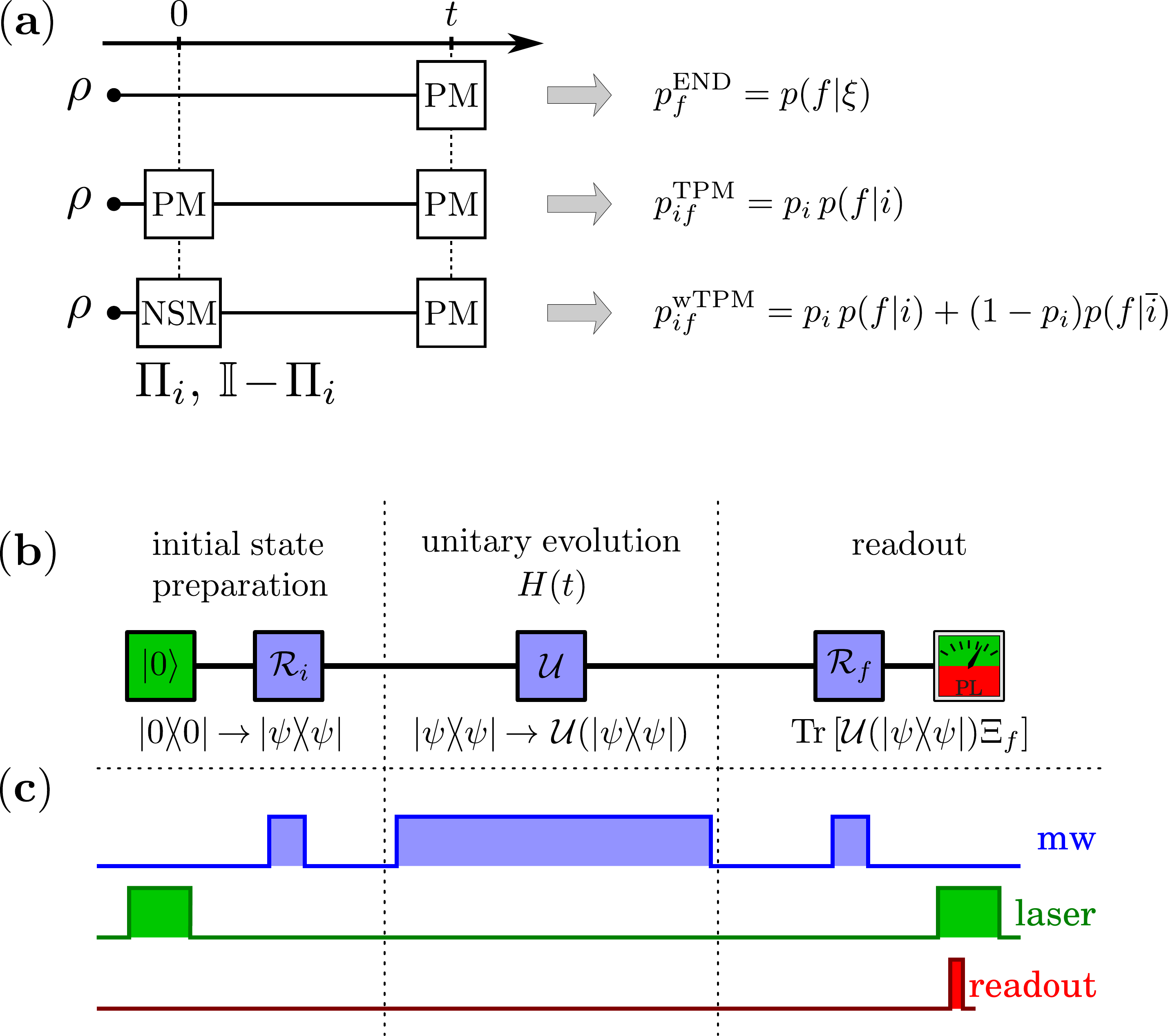}
\caption{
(a) Diagram showing the three different schemes END, TPM, and wTPM (from top to bottom) that are needed to access MHQ distribution. The schemes are based on projective measurements (PM) and non-selective measurements (NSM) of the Hamiltonian $H(t)$, our measurement observable. The END scheme consists of a single PM at time $t$, by directly initializing the system in $\rho=|\xi\rangle\!\langle\xi|$. Instead, the TPM scheme prescribes a PM at the beginning of the protocol, and a second PM at time $t$.  
Finally, the wTPM scheme is similar to the TPM, but the first PM is replaced by an NSM.
(b-c) Protocol for the measurement of the conditional probability $p(f|\psi)$ to get the energies $E_{f}(t)$ conditioned on the initial pure state $|\psi\rangle$. A resonant microwave (mw) [blue] is used to coherently control the spin. A green laser [green] is used to initialize the spin into $\ket{0}$, and to allow the optical read-out [red] of the probability for the system to be in the state $\ket{0}$. The mw gate $\mathcal{R}_i$~($\mathcal{R}_f$) maps $\ket{0}$ into $\ket{\psi}$ ($\Xi_f$ into $\ket{0}\!\!\bra{0}$).
}
\label{fig:protocol_sm}
\end{figure}

As discussed in the main text and in Ref.~\cite{companion_theory_paper}, the MHQ distribution $q_{if}$ can be reconstructed by combining three different measurement schemes: END, TPM, and wTPM. These schemes involve projective measurements and non-selective measurements, as illustrated in Fig.~\ref{fig:protocol_sm}a. 
These measurements provide the probability $p^{\mathrm{END}}_{f}$, and the joint probabilities $p^{\mathrm{TPM}}_{if}$ and $p^{\mathrm{wTPM}}_{if}$. 
In order to access such probabilities with our experimental platform, we take advantage of the fact that all $p^{\mathrm{END}}_{f}$, $p^{\mathrm{TPM}}_{if}$, $p^{\mathrm{wTPM}}_{if}$ can be expressed in terms of the conditional probabilities $p(f|\psi)$~[see Eq.~(9) in main text], 
as detailed in Fig.~\ref{fig:protocol_sm}a. 
This is convenient because, with our experimental setup, we can measure $p(f|\psi)$ for each of the different initial states $\ket{\xi}$, $\ket{i}$, and $\ket{\overline{i}}$, which allow us to reconstruct the END, TPM, and wTPM probabilities. In order to measure $p(f|\psi)$, we follow the protocol described in Fig.~\ref{fig:protocol_sm}b-c. 
The qutrit is initially prepared in the pure state $\ket{\psi}\in\{|\xi\rangle,|i\rangle,|\overline{i}\rangle\}$. To prepare $\ket{\psi}$, we initialize the system into $\ket{0}$ and we apply the mw gate $\mathcal{R}_i$ such that $\mathcal{R}_i(\ket{0}\!\!\bra{0}) = \ket{\psi}\!\!\bra{\psi}$. 
Note that this is possible since $|\psi\rangle$ is a pure state for any of the schemes (see main text). 
The qutrit then evolves unitarily under the Hamiltonian~[Eq.~(8) in main text] 
for a time $t$. During this unitary evolution, the quantum system exchanges work with the microwave field.
Finally, we measure the probability that the energy of the system is $E_f(t)$, i.e., we read-out $p(f|\psi)$. 
In order to achieve this, we apply a microwave gate $\mathcal{R}_f$ to the quantum system, such that $\mathcal{R}_f(\Xi_f)=\ket{0}\!\!\bra{0}$, and then we measure the PL intensity. 
As mentioned in the main text, the average PL value depends on the spin projection $m_S$, so we normalize it with respect to the PL reference levels to obtain the probability for the qutrit to be in the state $\ket{0}$: $\tr{\mathcal{R}_f(\mathcal{U}(\ket{\psi}\!\!\bra{\psi})) \ket{0}\!\!\bra{0}}={\rm Tr}[\mathcal{U}(\ket{\psi}\!\!\bra{\psi}) \mathcal{R}_f^{-1}(\ket{0}\!\!\bra{0})]=p(f|\psi)$. 
Note that the eigenstates of the Hamiltonian change in time, hence the gate $\mathcal{R}_f$ depends on the final time $t$. For a given initial state we perform independent experiments for several values of $t$ and for each of the three Hamiltonian projectors $\Xi_f(t)$. 
Due to the low photon collection efficacy and shot-noise of the detector, each of these experiments is repeated around $10^6$ times in order to obtain the average PL.

\section*{Role of non-classicality for enhanced extractable work}

In Fig.~3 in the main text we show the comparison between the work values $\langle w\rangle_t$ associated with the MHQ distribution and the average $\langle w\rangle_t^{\rm TPM}$ corresponding to the TPM scheme. In doing this, we also make use of the experimental data obtained from the implementation of the weak two-point measurement (wTPM) scheme.  

\begin{figure}
\centering
\includegraphics[width=0.45\columnwidth]{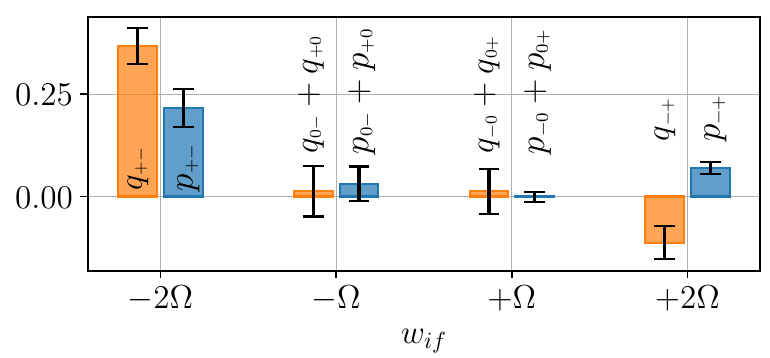} 
\caption{
Distributions $q_{if}$ and $p_{if}^{\rm TPM}$ ($p_{if}$ in labels to simplify notation) for each value of $w_{if}$ at $t=0.3$~$\mu$s, which is one of the times when the negativity $\aleph$ is maximized. The values for $w_{if}=0$ are not shown since they do not contribute to work extraction.
}
\label{fig:mean_workNew}
\end{figure}

From the figure, we can observe that the presence of non-classicality, in the form of negativity of the MHQs, entails a larger extractable work that is maximum when also $\aleph$ takes its maximum value. The physical explanation of this effects relies in the fact that non-classicality is able to transform the average work done by the system into extractable work, and vice versa. Let us analyze this statement more in detail. The average work $\langle w\rangle_t \equiv \sum_{i,f}q_{if}(E_f-E_i)$ of a MHQ work distribution can be equivalently written as
\begin{equation}
    \langle w\rangle_t = \sum_{i,f}\mu_{if}\|\v{q}\|{\rm sgn}(q_{if})\left(E_{f}(t) - E_{i}(0)\right),
\end{equation}
where the set of $\mu_{if} \equiv |q_{if}|/\|\v{q}\|$ forms a classical probability distribution respecting the Kolmogorov's axioms of probability theory, $\|\v{q}\| \equiv \sum_{if}|q_{if}|$ that is denoted as \emph{total negativity} ($\|\v{q}\|=1$ if no MHQs are negative), and ${\rm sgn}(\cdot)$ is the sign function that is equal to $+1$ if $(\cdot)$ is positive and $-1$ otherwise. Hence, by defining $\overline{E}_{f}(t) \equiv \|\v{q}\|{\rm sgn}(q_{if})E_{f}(t)$ and $\overline{E}_{i}(0) \equiv \|\v{q}\|{\rm sgn}(q_{if})E_{i}(0)$, we can interpret the average MHQ work $\langle w\rangle_t$ by means of a classical stochastic process defined by the set of probabilities $\{\mu_{if}\}$, i.e.,
\begin{equation}
\label{eq:average_W_classical_interpr}
    \langle w\rangle_t = \sum_{i,f}\mu_{if}\left(\overline{E}_{f}(t) - \overline{E}_{i}(0)\right).
\end{equation}
One can thus compare Eq.~(\ref{eq:average_W_classical_interpr}) with the average TPM work $\langle w\rangle_t^{\rm TPM}$, and then explain why non-classicality (negativity of MHQs in our experimental case-study) can entail a larger amount of extractable work.
In fact, when negativity of the single term $q_{if}$ is present, positive work terms $E_{f}(t) - E_{i}(0) \geq 0$ (corresponding to work done by the system) changes sign and they transforms in $\overline{E}_{f}(t) - \overline{E}_{i}(0) < 0$, i.e., work that can be extracted from the system. Moreover, always in case of negativity, the effective energies $\overline{E}$ are obtained by multiplying $E$ for $\|\v{q}\| \geq 1$. Hence, the extractable work originated by non-classicality is larger --in absolute value-- than the corresponding positive work terms (done by the system) that enter the TPM work average. Fig.~\ref{fig:mean_workNew} shows an instance of this key aspect by comparing the TPM and MHQ distributions at a time instant corresponding to one of the peaks of the negativity of MHQs in our experiment. All the negativity in the MHQ work distribution is associated with the largest exciting transition, $w_{-+} = 2\Omega$ that, classically, would contribute to the work done but quantumly it enhances the work extraction. This negativity is destroyed in the TPM scheme, resulting in decreased work extraction. 

It worth noting that the fact that the value of the extracted work according to the MHQ is larger than the one evaluated with the TPM scheme \textit{per se} is not a proof of non-classicality. In fact, let us consider the case entailed by our experiment with a unitary work protocol $U$, an initial pure state $\ket{\xi}$, and rank-1 projectors on the initial and final energy eigenstates, i.e., $\Pi_i=|E_i(0)\rangle\!\langle E_i(0)|$ and $\Xi_f=|E_{f}(t)\rangle\!\langle E_{f}(t)|$. Then, the MHQ is given by
\begin{align}
    q_{if}=\Re{\bra{\xi}E_i(0)\rangle\bra{E_i(0)}U^\dag\ket{E_f(t)}\bra{E_f(t)}U\ket{\xi}},
\end{align}
with 
\begin{align}
    & \bra{\xi}E_i\rangle=e^{-i\Phi^{(I)}_i}\sqrt{p_i}\\
    & \bra{E_i}U^\dag\ket{E_f(t)}=e^{-i\Phi^{(C)}_{if}}\sqrt{p(f|i)}\\
    & \bra{E_f(t)}U\ket{\xi}=e^{-i\Phi^{(E)}_{f}}\sqrt{p^{\rm END}_{f}}\,.
\end{align}
Here, $p_i \equiv |\bra{E_i(0)}\xi\rangle|^2$ are the initial probabilities, $p(f|i) \equiv |\bra{E_f(t)}U\ket{E_i(0)}|^2$ are the conditional probabilities of observing the outcome $E_f(t)$ in the final projective measurement conditioned on having as initial state $\ket{E_i(0)}$, 
and $p^{\rm END}_{f}=|\bra{E_f(t)}U\ket{\xi}|^2$ are the probabilities of the end-point measurement protocol. Thus, noticing that the TPM scheme probabilities are given by $p^{\mathrm{TPM}}_{if} = p_{i}\,p(f|i)$, we end up with 
\begin{align}
    q_{if} = A_{if}\sqrt{p^{\rm TPM}_{if}p^{\rm END}_f}\,,
\end{align}
where we have defined $A_{if} \equiv \cos{\left(\Phi^{(I)}_i+\Phi^{(C)}_{if}+\Phi^{(E)}_f\right)} \in[-1,1]$. From the last expression one can see that the MHQ can be larger or smaller than their TPM counterpart depending on the end-point probabilities and the \textit{activities} $A_{if}$. The latter encode quantum interference and, indeed, $q_{if}\geq 0 \Longleftrightarrow A_{if}\geq 0$. Now, let us assume that the probabilities of the TPM and end-point schemes are fixed, and define the extracted (when positive) work as 
\begin{equation}
    W_{\rm ext}\equiv -\langle w\rangle_t =\bra{\xi}H(0)\ket{\xi}-\bra{\xi}U^\dag H(t)U\ket{\xi}\,.
\end{equation}
Hence, we see immediately that, if $A_{if}\geq 0$ $\forall\,i,f$, then 
\begin{equation}
    W_{\rm ext}\leq \sum_{i,f \,\, \text{s.t.} \,\, E_i>E_f}(E_i - E_f)\sqrt{p^{\rm TPM}_{if}p^{\rm END}_f}\,,
\end{equation}
which gives a \textit{classical} upper bound to the extracted work.
Fig.~3 in the main text shows that, in our experiment, we violate this classical bound. Such violation is a witness of non-classicality of the dynamical process of work.
An analogous bound can be found when considering the absorbed work: $\langle w\rangle_t \leq \sum_{i,f \,\, \text{s.t.} \,\, E_i<E_f}(E_f - E_i)\sqrt{p^{\rm TPM}_{if}p^{\rm END}_f}\,$.

\section*{Maximizing the amount of extractable work: Numerical analysis for generic parameters of the NV center Hamiltonian}

As described in the main text, the interaction between an NV center with an on-resonance microwave field results in the Hamiltonian~[Eq.~(8) in the main text] (expressed in the MW rotating frame) that, after the rotating wave approximation, reads as
\begin{equation}
H(t) = \Omega_1\left(S_{x1}\cos\phi_1 t +S_{y1}\sin\phi_1 t \right) +  \Omega_2\left( S_{x2}\cos\phi_2 t -S_{y2}\sin\phi_2 t  \right). 
\label{eq:Hamiltonian_suppl}
\end{equation}

Here, we aim to identify {\it (i)} the parameters for which $q_{if}$ is minimized (for at least one set of $i,f$), {\it (ii)} the parameters for which the average MHQ extracted work $W_{\rm ext}=-\langle w\rangle_t$ is maximized, and {\it (iii)} the parameters for which the negativity $\aleph$ is maximized. Note that, a single set of parameters do not necessarily fulfill all the above conditions {\it (i)}, {\it (ii)}, {\it (iii)}. In order to achieve this, we run numerical simulations for 10000 different sets of random parameters $\{\Omega_1^\mathrm{R},\phi_1^\mathrm{R},\Omega_2^\mathrm{R},\phi_2^\mathrm{R}\}$, such that $\{\Omega_1,\phi_1,\Omega_2,\phi_2\} = \{\Omega_1^\mathrm{R},\phi_1^\mathrm{R},\Omega_2^\mathrm{R},\phi_2^\mathrm{R}\}$, and we calculate $\min[q_{if}]$, $\min[\langle w\rangle_t]$, and $\max[\aleph]$ for each set of parameters. The $\min[\cdot]$ and $\max[\cdot]$ are calculated over the time interval $t \in \big(0,2\pi/\sqrt{2(\Omega_1^2+\Omega_2^2) + \phi_1^2}\big)$. The longest time value in this interval corresponds to a period of the dynamics in the case with $\phi_2=\phi_1$. In addition, the considered random parameters correspond to random floating-point numbers in the intervals $\Omega_1^\mathrm{R} \in [1,20]$~MHz, $\phi_1^\mathrm{R} \in  [-2\Omega_1^\mathrm{R},2\Omega_1^\mathrm{R}]$, $ \Omega_2^\mathrm{R} \in [1,20]$~MHz, and $ \phi_2^\mathrm{R}\in [-2\Omega_2^\mathrm{R},2\Omega_2^\mathrm{R}]$. 
For each set of parameters, the initial state is a random pure state $\rho = \ket{\xi}\!\!\bra{\xi}$, with $\ket{\xi} = a e^{i \phi_a}\ket{+1} + b e^{i \phi_b}\ket{0} + \sqrt{1-a^2-b^2}\ket{-1}$, where $a$, $b$, $\phi_a$, $\phi_b$ are random floating-point numbers in the intervals $[0,1]$, $[0,\sqrt{1-a^2}]$, $[0,2\pi)$, and $[0,2\pi)$ respectively. 

\begin{figure}[h!]
\centering
\includegraphics[width=0.925\columnwidth]{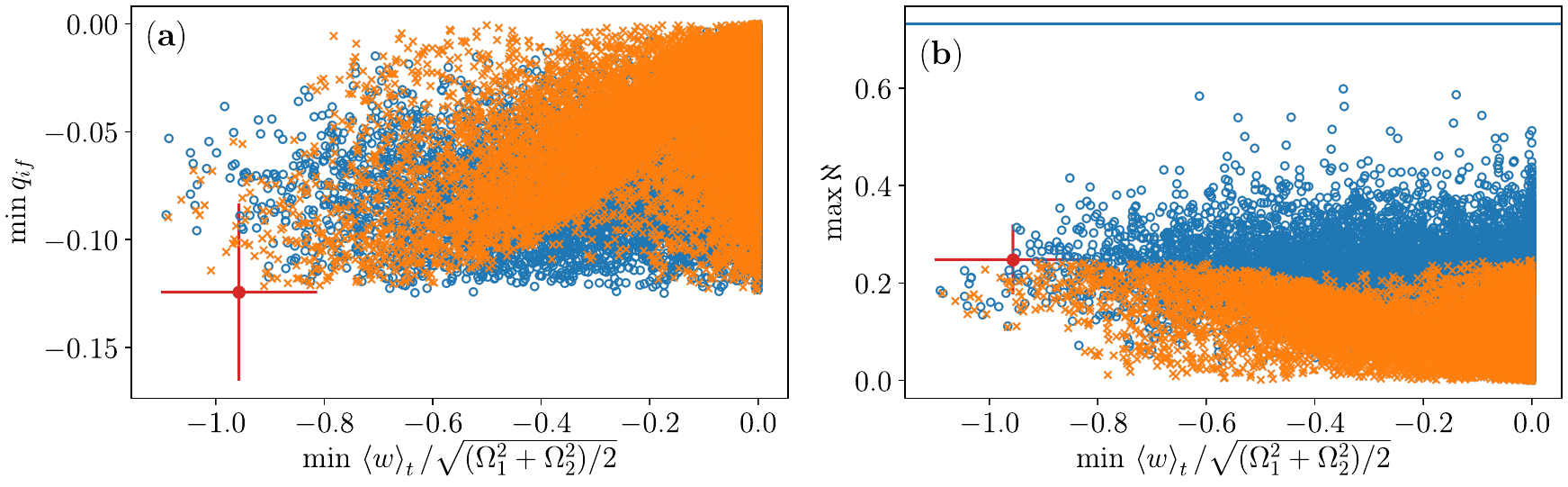}
\caption{
Results of the numerical simulations. 
In both panels (a)-(b), each empty blue circle represents the result for a set of random parameters of the system Hamiltonian and a random initial pure state. 
For each empty blue circle there are two orange crosses corresponding to the cases $\phi_1=\phi_2=\phi_1^\mathrm{R}$ and $\phi_1=\phi_2=\phi_2^\mathrm{R}$.
Instead, the red circle with the error bars represents the experimentally measured values, as detailed in the main text. 
Finally, in panel (b), the horizontal line denotes the upper bound of the non-classicality measure $\aleph$. Such a bound is equal to $\sqrt{d}-1$, where $d=3$ is the dimension of the Hilbert space of the system~\cite{companion_theory_paper}.
}
\label{fig:numeric_simul}
\end{figure}

The results of the numerical simulations are summarized in Fig.~\ref{fig:numeric_simul}. 
From these results it is evident that the condition $\phi_1=\phi_2$ allows for the minimization of $q_{if}$, as well as the maximization of $W_{\rm ext}$. In contrast, in order to maximize the negativity $\aleph$, it is more convenient to select the parameters of the system Hamiltonian such that $\phi_1\neq\phi_2$.

\end{document}